# Exploration in Booster Reuse Employing Aerosurfaces

Timur Dayanov

A new idea for reusing rockets, based on the Energia 2 launch vehicle, using deployable wings is proposed. A mission design is planned with an expected maximum altitude of 500 m. Following that systems engineering was used to manage the design process of separate components, which in turn led to the manufacturing plans for the separate components. The sizing studies were used to determine how big each component would need to be, and a weight and balances sheet was used to determine the position of the wings for optimal static margin and stability percentage. A derivation was made using basic Newtonian mechanics for determining the maximum mass of the rocket given a target altitude. Following that Computer Assisted Design (CAD) drawings were created of critical components before outlining the necessary structural tests and the flight test plan. A MatLab script to determine static stability was used which confirmed the stability of the booster at angle of attacks from 0 to 5 degrees. A MatLab script to determine glide ratio and range was used on the booster and Energia 2 to show scalability, and which was found feasible.

I. INTRODUCTION

A major problem and constraint in rocketry is cost. One of the ways to reduce the cost per launch is to reuse the rocket, so a new reuse system was designed and selected for its minimum weight added and inexpensive construction. The system presented is inspired by the plans for the Russian Energia 2, that has side boosters which deploy wings and make a controlled descent to the Earth [1].

Prior to beginning the design process, research was done on previous reusability systems, including the Falcon 9, the Electron, and the proposed plans for Energia 2 side boosters. Falcon 9 saves fuel to perform a landing burn, the Electron is caught by its parachute in the air, and the Energia 2 side boosters deploy wings and land on a runway [1,2,3].

II. MISSION DESIGN

The purpose of the mission is to test the concept of a deployable wing system with minimal extra components. The trajectory of the rocket will start at a 5-degree angle, then transition to horizontal flight at the peak, level out after gaining speed, and glide at a steady rate of descent to the ground. The mission's phases are listed below in Table 1.

TABLE 1. FLIGHT PLAN

| Mission Phase | Altitude (m) | Velocity (m/s) | Time (s) | Success Metric |
|---|---|---|---|---|
| 1: Liftoff | 0 | 0 | 0 | • Engine Ignites<br>• Rocket Lifts Off<br>• Rocket Stays Upright |
| 2: Engine Cutoff | 121 | 86.4 | 2.8 | • Rocket Reaches Engine Cutoff<br>• Engine Burns Out |
| 3: Apex[a] | 500 | 0 | 11.6 | • Rocket Reaches Apex<br>• Rocket Deploys Wings<br>• Rocket Goes Horizontal |
| 4: Landing | 0 | 1[b] | varies | • Rocket Reaches Half Maximum Achieved Height<br>• Rocket Reaches Land<br>• Rocket Fuselage Stays Intact<br>• Rocket Wings Stay Intact |

a. includes transition to horizontal flight
b. on y axis

III. SYSTEMS ENGINEERING

Systems engineering is used to manage the design process and integration of separate individual design components. A list of rocket parts with their design requirements and necessary components is listed below in Table 2 and contains the systems and related subsystems. The purpose of each subsystem is described after it, as it assists in the design process. It is important to note that most of the research is focused on the reusability system, and the other subsystems are primarily used in mass calculations.

TABLE 2. ROCKET PARTS

| Subsystem | Description | Rocket Parts |
|---|---|---|
| *First Stage* | | |
| Engine | Provides thrust to the rocket and is responsible for propelling the rocket to the desired altitude | • Solid Fuel Tank<br>• Ignition |
| Fuselage | Maintains structural integrity, keeps everything together, and acts as housing and an aerodynamic surface | • Engine Attachment<br>• Brain Housing<br>• Housing for Wing System<br>• Fin Attachment<br>• Fuel Tank Attachment<br>• Structure (material, shape) |
| Guidance/ Brain | Tracks the rocket path, make the decision to stage, and controls aerosurfaces on descent | • Sensors for telemetry<br>• CPU<br>• Wiring<br>• Radio<br>• Servos |
| Fins | Keep rocket vertical in powered flight, allow it to maintain heading during gliding, stabilizers | • Attachment to fuselage<br>• Aerodynamic shape |
| *Reusability System* | | |

| Subsystem | Description | Rocket Parts |
|---|---|---|
| Wing System | Deploys at apex, transitions between vertical and horizontal flight, maintains heading when not controlled, generates lift, and houses control surfaces | - Deployment axis (attached to servo)<br>- Elevons (attached to servos)<br>- Wing shape (generates lift)<br>- Connection to Housing |

*Manufacturing Plans*

A "Hi-Tech" rocket from Apogee Rockets will be purchased assembled [4]. It will be modified to hold two servos and an aerodynamic cover in order to hold and deploy the wings without significantly increasing drag. significantly. The servos will be placed as close to the fuselage as possible to minimize the shift in the center of mass, and will be attached by removing material, slotting the servos in, and securing them from falling out. The wing cover will be screwed onto the fuselage by its structure. The wings themselves will be manufactured by using wood ribs and stiff spar to create the frame. A foil would then be pulled over the structure. This approach minimizes weight which is the primary concern of this system.

*Preliminary Sizing Studies*

Sizing is important in calculating the mass, volume, and center of mass, which are needed to understand how the rocket will fly under given conditions. Using a weights and balances sheet, the rocket is configured to have the optimal positioning of the wings relative to the fuselage, so as to provide enough lift to safely descend to the ground. Since the scale is based off of the "Hi-Tech" rocket, the wings cannot be any longer than the length from their rotating point and no wider than the width of the rocket. The mass of the rocket is determined, using a calculator created in Desmos (https://www.desmos.com/calculator/go5zuzqbbd) which is set to find the maximum weight given specific engine capabilities and the dry mass of the original rocket. The formula derivation that relates thrust, mass, altitude, and burn time is shown below (Fig. 1).

$$v_0 = 0 m/s$$
$$y_0 = 0 m$$
$$t_0 = 0 s$$
$$a_{p1} = \frac{F}{m_{max}} - 9.8 m/s^2$$
$$a_{p2} = -9.8 m/s^2$$

$$\Delta y = \frac{1}{2}(a)(t^2) + v_0(t)$$
$$v^2 = (v_0)^2 + 2a\Delta y$$

$$v_1 = a_{p1}(t_1)$$
$$t_1 = t_{burn}$$
$$y_1 = \frac{1}{2}(a_{p1})(t_1)^2$$

$$y_2 = y_1 - \frac{v_1^2}{2a_{p2}} = y_{max}$$
$$y_{max} = \frac{1}{2}(\frac{F}{m_{max}} - 9.8 m/s^2)(t_{burn})^2 + \frac{((\frac{F}{m_{max}} - 9.8 m/s^2)t_{burn})^2}{19.6 m/s^2}$$
$$y_{max} = \frac{F}{2m}(t_1)^2 - \frac{g}{2}(t_1)^2 + \frac{(\frac{F}{m}t_1 - g(t_1))^2}{2g}$$
$$y_{max} = \frac{F}{2m}(t_1)^2 - \frac{g}{2}(t_1)^2 + \frac{F^2}{2m^2(g)}(t_1)^2 - \frac{F}{m}(t_1)^2 + \frac{g}{2}(t_1)^2$$
$$y_{max} = \frac{1}{m}(\frac{F}{2}(t_1)^2 - F(t_1)^2) + \frac{1}{m^2}(\frac{F^2}{2g}(t_1)^2)$$
$$m^2(y_2) = m(\frac{F}{2}(t_1)^2 - F(t_1)^2) + (\frac{F^2}{2g}(t_1)^2)$$
$$m = \frac{(\frac{F}{2}(t_1)^2 - F(t_1)^2) \pm \sqrt{(\frac{F}{2}(t_1)^2 - F(t_1)^2)^2 - 4(y_2)(\frac{F^2}{2g}(t_1)^2)}}{2y_2}$$

Figure 1. Mass calculation formula.

The servos are placed closer to the sides of the fuselage to allow for a wider wing span. If the servos had been placed next to each other, the wingspan would have been the length of both wings added together. However, the farther apart the servos are placed, the greater the wingspan becomes because it is the length of both wings added together plus the space between the servos.

The weights and balances sheet (shown below in Table 3) is used to find the center of gravity (CG), aerodynamic chord (AC) and the static margin and stability percentage (Table 4). It is important for the thrust to line up with the neutral axis. The neutral axis is determined to be the center of the fuselage because a cylinder has equal loads from all sides when flying against the force of gravity (shown in Section IV).

The deployed configuration takes into account a horizontal rocket acting as a plane, so it is important to calculate the different center of gravity. Since the center of the wings' mass is farther back and the static margin is negative, it creates a tendency to maintain a slightly nose down position in order to reach the ground while expending minimum energy.

TABLE 3. WEIGHT AND BALANCES

| Component | X Length (cm) | Weight (g) | X distance from (0.0) to back of component | X Coord (cm) | Y Coord (cm) |
|---|---|---|---|---|---|
| **Configuration: Stowed** | | | | | |
| Fuselage | 103.5 | 512.8 | 0 | 51.8 | 0 |
| Engine | 12.7 | 197 | 0 | 6.4 | 0 |
| Wing Housing | 70 | 179 | 31 | 66 | 6.7 |
| Wings | 68 | 23.8 | 35 | 69 | 6.7 |
| Nose Cone | 22.9 | 56.5 | 103.5 | 115 | 0 |
| Fins (3x) | 15.2 | 54.4 | 2 | 9.6 | 0 |
| Total | | 1023.5 | | | |
| **Configuration: Deployed** | | | | | |
| Fuselage | 103.5 | 512.8 | 0 | 51.8 | 0 |
| Engine | 12.7 | 197 | 0 | 6.4 | 0 |
| Wing Housing | 70 | 179 | 31 | 66 | 6.7 |
| Wings | 5 | 23.8 | 35 | 37.5 | 6.7 |
| Nose Cone | 22.9 | 56.5 | 103.5 | 115 | 0 |
| Fins (3x) | 15.2 | 54.4 | 2 | 9.6 | 0 |
| Total | | 1023.5 | | | |

TABLE 4. CENTER OF GRAVITY

| **Configuration: Stowed** | |
|---|---|
| X CG Location (cm) | 47.19079629 |
| Y CG Location (cm) | 1.327562286 |

| Configuration: Deployed | |
|---|---|
| X CG Location (cm) | 46.45830972 |
| Y CG Location (cm) | 1.327562286 |
| AC Location (cm) | 38.75 |
| Static Margin | -7.708309722 |
| Stability (%) | -7.447642243 |
| Total | |

The bill of materials is listed below in Table 5.

TABLE 5. BILL OF MATERIALS

| Subsystem | Materials |
|---|---|
| Engine | • Cesaroni P54 I150 |
| Fuselage | • Hi-Tech (Model 07659) |
| Wing System | • balsa wood<br>• 2 2x1 Sheets<br>• 2 servos for deploying wings<br>• 2 servos for air surface control |
| Fuel Tank | • part of the engine |
| Guidance/Brain | • Raspberry PI/Arduino |
| Fins | • wood cutouts from 3 1x1/2 ft |

A Computer Assisted Design (CAD) drawing of the fuselage and nose cone is shown below (Fig. 2).

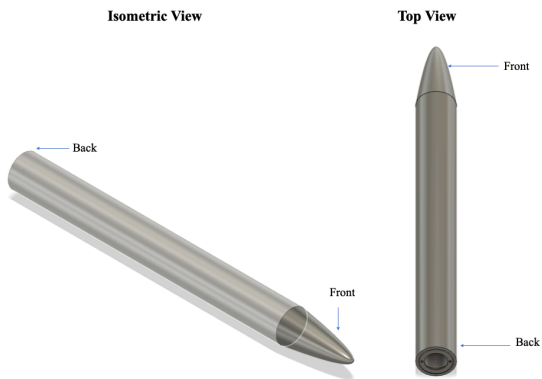

Figure 2. Isometric and Top View of Fuselage, CAD.

A CAD drawing of wing (using the airfoil NACA 2414 shown in Fig. 3) and wing housing is shown below (Fig. 4).

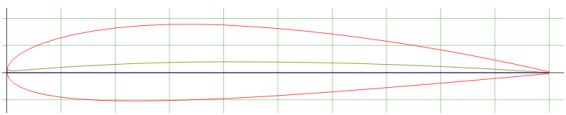

Figure 3. Air Foil for Wing, NACA 2414.

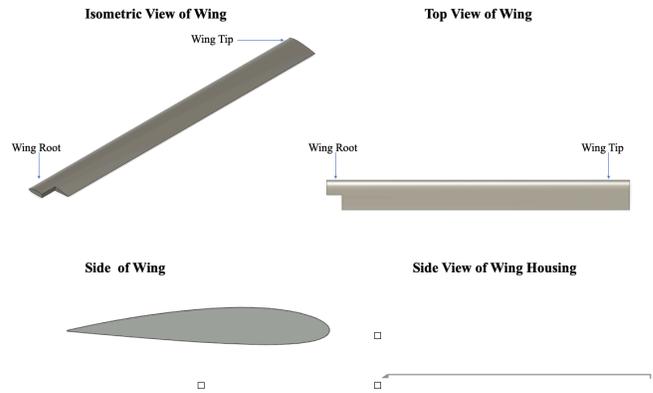

Figure 4. Wing Shape and Wing Housing, CAD

The fins from Hi-Tech rocket are shown below (Fig. 5).

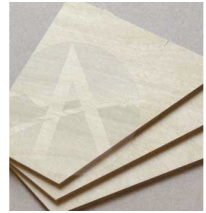

Figure 5: Standard fins with the Hi-Tech rocket.

IV. PLANNED STRUCTURAL TESTS

The test plans necessary to verify that the rocket is fit for flight are listed below with the specific part being tested, assumptions, and representation of in-flight circumstances.

*A. FUSELAGE*

Test Article: 2x3/4 ft^2 Aluminum Cylinder

Compression Test: Represents the maximum G forces that the fuselage experiences during ascent. It assumes maximum G forces when the rocket is completely vertical. The test is depicted in Fig. 6.

5 G Load.

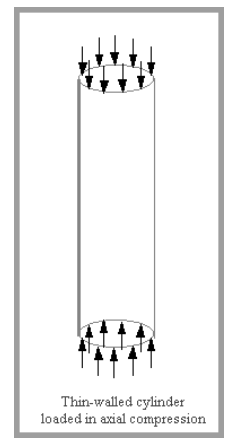

Figure 6. Visual depiction of compression test.

Worst Case Bend Test: Represents G forces if fuselage flips over. Assumes air drag exerts 2 Gs at top speed. The test is depicted in Fig.7.

3 G load.

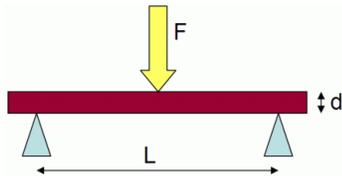

Figure 7. Visual depiction of worst case bend test.

### B. WINGS

Test Article: 2 ft. long x 1/2 ft. wide aluminum pieces, 2 in. thick

Wing Tip Deflection Test: Represents the lift force that the wings must sustain to keep the rocket afloat. It assumes that speed does not affect the stress that the wings endure and that forces only act vertically. The test is depicted in Fig. 8.

1.5 G load for each wing.

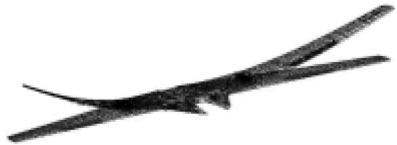

Figure 8. Visual depiction of wing tip deflection test.

## V. STABILITY

To give a more detailed look into stability, the airfoil was analyzed in XFoil to obtain coefficients at different angles of attack. The MatLab script from "Determine Nonlinear Dynamics and Static Stability of Fixed-Wing Aircraft" was used with the values obtained from XFoil shown in Fig. 9 and Table 6 and printed out where the rocket would be statically stable, shown in Table 7 [5].

Static stability refers to an aircraft's response to a perturbation that causes an oscillation when it's flying at a flight condition parameter. If the aircraft's oscillation dampens, then it's statically stable. If the oscillation doesn't dampen or grow, then it's neutrally stable. If the oscillation grows, then it's statically unstable.

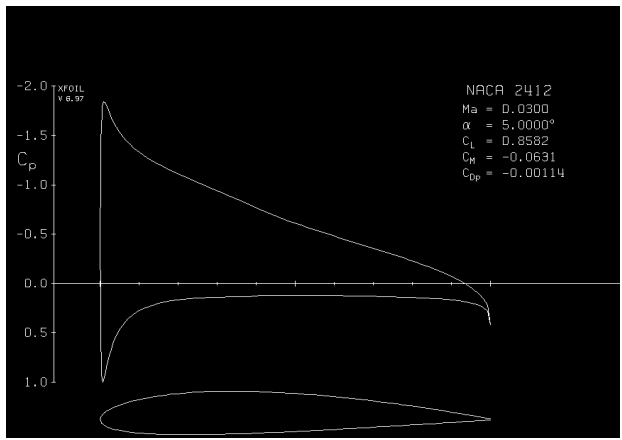

Figure 9. Airfoil results from XFoil.

TABLE 6. COEFFICIENTS FROM XFOIL

|    | Angle of Attack | Value  |
|----|-----------------|--------|
| CD | Zero            | -0.015 |
| CL | Zero            | 0.256  |
| Cm | Zero            | -0.056 |
| CD | Alpha           | -0.018 |
| CL | Alpha           | 1.456  |
| Cm | Alpha           | -0.071 |

TABLE 7. STATIC STABILITY AT ALPHA

|    | U        | V        | W      | Alpha  | Beta     | P       | Q       | R       |
|----|----------|----------|--------|--------|----------|---------|---------|---------|
| FX | Unstable |          |        |        |          |         |         |         |
| FY |          | Unstable |        |        |          |         |         |         |
| FZ |          |          | Stable |        |          |         |         |         |
| L  |          |          |        |        | Unstable | Neutral |         |         |
| M  | Unstable |          |        | Stable |          |         | Neutral |         |
| N  |          |          |        |        | Neutral  |         |         | Neutral |

The only coefficients filled in are the at the zero angle of attack and an arbitrary alpha angle of attack, as those are all that's needed, for determining static stability in the pitch axis.

## VI. GLIDE RATIO AND RANGE CALCULATION

The MatLab script "Calculating Best Glide Qualities" was used to calculate the glide ratio, which is important in calculating the range of the booster [6].

Based on the graph in Fig. 10 the point where the glide ratio is the highest is when the air speed is 48.1 knots, with the glide ratio 20.7, giving a range of 10,350 m. The best glide angle of attack was calculated to be -2.7 degrees.

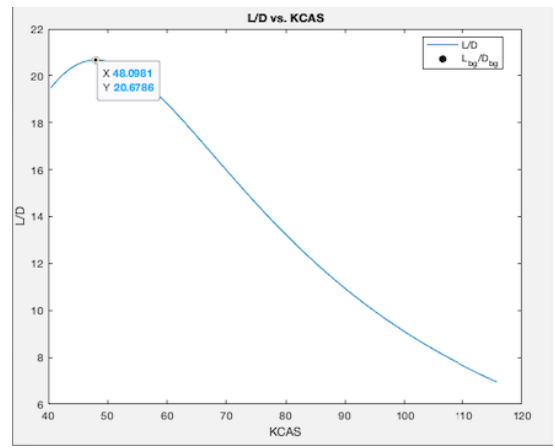

Figure 10. Glide ratio calculation.

To show how the design and procedure can be scaled, the same script was ran on the specifications of the Energia 2 booster, with a result of a glide ratio of 14.6, which at the altitudes that the Energia 2 booster operates at, gives a 438 km range shown in Fig. 11 assuming best glide conditions are met, which are an angle of attack -3.9 degrees and an airspeed of 374 knots.

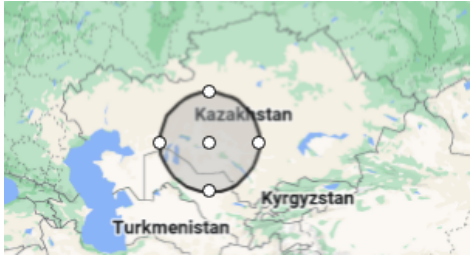

Figure 11. Range around Baikonur Cosmodrome for Energia 2

## VII. FLIGHT TEST PLAN

The flight test would be performed in a designated model rocketry launch zone. The key milestones to achieve in order to validate the concept and its safety are the apex where the rocket should deploy its wings and the landing. Horizontal velocity is needed in order to generate initial lift, so the rocket will have to perform a slight nose down before leveling out. It is critical that the rocket achieves a stable velocity before reaching 400 m. in altitude in order to guarantee the safety of observers as well as its own integrity. After that, it can glide down to a designated landing site. The main concern with this flight plan is that the rocket won't accumulate enough horizontal airspeed, but that would not be a problem with larger scale rockets, as at booster separation the horizontal velocity of a typical first stage is more than enough.

The Energia's flight plan is one that meets these criteria, as shown in Fig. 12. The rocket ascends to 54 km, where the booster separates, and it reaches a maximum height of 75 km before deploying wings. The booster deploys wings at 18 km, at a speed of 250 m/s, which decreases the Mach number from 1.1 to 0.75, and lands at a speed of 83 m/s [1].

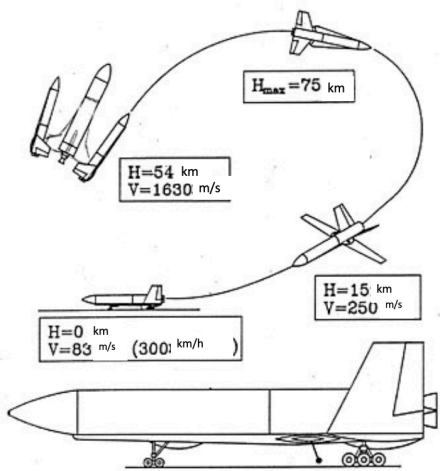

Figure 12. Energia 2 flight plan [1]

## VIII. CONCLUSION

In this work, a novel approach to reusable rocketry is presented, a mission profile is developed, a design is pitched with planned test verifications that include a compression test, worst case bend test, and a wing tip deflection test. The design used a NACA 2412 airfoil, which was drawn in CAD, and had a weight of 1.0235 kg, length of 1.04 m, and wing span of 1.4 m when deployed. The design specification is inputted into MatLab scripts to determine the static stability, which is stable at 0 and 5 degrees of pitch, and the range of the rocket, which at an angle of attack of -2.7 degrees, an airspeed of 48.1 knots, is 10,350 m. Future work would include the manufacture of a prototype, the execution of planned structural tests, and a flight test. To evaluate scalability, the glide range calculations were performed on the Energia 2 side booster [1]. With an angle of attack of -3.9 degrees, and an airspeed of 374 knots the best range of 438 km is achieved. As such the scalability of this design is feasible, as it also resolves the main concern that the rocket will not reach a great enough horizontal velocity relative to the ground. Rockets that perform a gravity turn have a greater horizontal velocity, leading to a more effective wing system. Potential scaling of the design would include modifying orbital class launch vehicles to implement such a recovery system.

### ACKNOWLEDGMENT

I would like to express my gratitude to my mentors, Humberto Caldelas, PhD Candidate at MIT, and Hannah Stroud, a PhD student at Texas A&M, for their support and guidance.

## APPENDIX A: MATLAB CODES

### A. Static Stability of Model Rocket

```
elevator = fixedWingSurface("Elevator", "on","Symmetric",[-20,20])
elevator.Coefficients = fixedWingCoefficient("Elevator")
aileron = fixedWingSurface("Aileron", "on", "Asymmetric", [-20,20], ...
    "Coefficients", fixedWingCoefficient("Aileron"));
wing = fixedWingSurface("Wing","Surfaces", aileron);
RocketProperties = Aero.Aircraft.Properties(...
    "Name"       , "Rocket", ...
    "Type"       , "General Aviation", ...
    "Version"    , "1.0", ...
    "Description", "Rocket Thing")
BodyCoefficients = {
   'CD', 'Zero', -0.015;
   'CL', 'Zero', 0.256;
   'Cm', 'Zero', -0.056;
   'CD', 'Alpha', -0.018;
   'CL', 'Alpha', 1.456;
   'Cm', 'Alpha', -0.071;
   };
Rocket = Aero.FixedWing(...
```

```
"Properties"     , RocketProperties, ...
"UnitSystem"     , "Metric", ...
"AngleSystem"    , "Radians", ...
"TemperatureSystem", "Celsius", ...
"ReferenceArea"  , 0.98, ...
"ReferenceSpan"  , 1.4, ...
"ReferenceLength", 1.04, ...
"Surfaces"       , [wing])
Rocket = setCoefficient(Rocket, BodyCoefficients(:, 1),
BodyCoefficients(:, 2),
BodyCoefficients(:, 3));
CruiseState = Aero.FixedWing.State(...
    "UnitSystem",Rocket.UnitSystem,...
    "AngleSystem",Rocket.AngleSystem, ...
    "TemperatureSystem",Rocket.TemperatureSystem, ...
    "Mass", 1.0235, ...
    "U", 50, ...
    "AltitudeMSL",500);
CruiseState.Inertia.Variables = [
    1, 0, 0;
    0, 1, 0;
    0, 0, 1;
];
CruiseState.CenterOfGravity = [0.456, 0 , 0] .* Rocket.ReferenceLength;
CruiseState.CenterOfPressure = [0.25, 0, 0] .* Rocket.ReferenceLength;
CruiseState.Environment = aircraftEnvironment(Rocket,"ISA",CruiseState.AltitudeMSL);
RocketState = setupControlStates(RocketState, Rocket)
staticStability(Rocket, RocketState)
```

### *B. Glide Ratio for Model Rocket:*

```
W = 2.2; % weight, lbf
S = 0.377;  % wing reference area, ft^2;
A = 14; % wing aspect ratio
C_D0 = 0.018; % flaps up parasite drag coefficient
e = 0.70; % airplane efficiency factor
h = 1640; % altitude, ft
phi = 0; % bank angle, deg
h_m = convlength(h,'ft','m');
[T, a, P, rho] = atmoscoesa(h_m, 'Warning');
rho = convdensity(rho,'kg/m^3','slug/ft^3');
TAS_bg = sqrt( (2*W)/(rho*S) )...
 *( 1./( (4*C_D0.^2) + (C_D0.*pi*e*A*(cos(phi)^2)) )).^(1/4); % TAS, fps
KTAS_bg = convvel(TAS_bg,'ft/s','kts')';
KCAS_bg = correctairspeed(KTAS_bg,a,P,'TAS','CAS')';
gamma_bg_rad = asin( -sqrt((4.*C_D0')./(pi*e*A*cos(phi)^2 + 4.*C_D0')) );
gamma_bg = convang(gamma_bg_rad,'rad','deg');
D_bg = -W*sin(gamma_bg_rad);
L_bg =  W*cos(gamma_bg_rad);
qbar = dpressure([TAS_bg' zeros(size(TAS_bg,2),2)], rho);
C_D_bg = D_bg./(qbar*S);
C_L_bg = L_bg./(qbar*S);
TAS = (70:200)'; % true airspeed, fps
KTAS = convvel(TAS,'ft/s','kts')'; % true airspeed, kts
KCAS = correctairspeed(KTAS,a,P,'TAS','CAS')'; % corrected airspeed, kts
qbar = dpressure([TAS zeros(size(TAS,1),2)], rho);
Dp = qbar*S.*C_D0;
Di = (2*W^2)/(rho*S*pi*e*A).*(TAS.^-2);
D = Dp + Di;
L = W;
h1 = figure;
plot(KCAS,L./D);
title('L/D vs. KCAS');
xlabel('KCAS'); ylabel('L/D');
hold on
plot(KCAS_bg,L_bg/D_bg,'Marker','o','MarkerFaceColor','black',...
    'MarkerEdgeColor','black','Color','white');
hold off
legend('L/D','L_{bg}/D_{bg}','Location','Best');
h2 = figure;
plot(KCAS,Dp,KCAS,Di,KCAS,D);
title('Parasite, induced, and total drag curves');
xlabel('KCAS'); ylabel('Drag, lbf');
hold on
plot(KCAS_bg,D_bg,'Marker','o','MarkerFaceColor','black',...
    'MarkerEdgeColor','black','Color','white');
hold off
legend('Parasite, D_p','Induced, D_i','Total, D','D_{bg}','Location','Best');
```

### *C. Glide Ratio for Energia 2.*

```
W = 750000; % weight, lbf
S = 2000;  % wing reference area, ft^2;
A = 14; % wing aspect ratio
C_D0 = 0.037; % flaps up parasite drag coefficient
e = 0.72; % airplane efficiency factor
h = 98425; % altitude, ft
phi = 0; % bank angle, deg
h_m = convlength(h,'ft','m');
[T, a, P, rho] = atmoscoesa(h_m, 'Warning');
rho = convdensity(rho,'kg/m^3','slug/ft^3');
TAS_bg = sqrt( (2*W)/(rho*S) )...
 *( 1./( (4*C_D0.^2) + (C_D0.*pi*e*A*(cos(phi)^2)) )).^(1/4); % TAS, fps
KTAS_bg = convvel(TAS_bg,'ft/s','kts')'
KCAS_bg = correctairspeed(KTAS_bg,a,P,'TAS','CAS')';
gamma_bg_rad = asin( -sqrt((4.*C_D0')./(pi*e*A*cos(phi)^2 + 4.*C_D0')) );
gamma_bg = convang(gamma_bg_rad,'rad','deg');
D_bg = -W*sin(gamma_bg_rad);
L_bg =  W*cos(gamma_bg_rad);
qbar = dpressure([TAS_bg' zeros(size(TAS_bg,2),2)], rho);
C_D_bg = D_bg./(qbar*S);
C_L_bg = L_bg./(qbar*S);
TAS = (70:200)'; % true airspeed, fps
KTAS = convvel(TAS,'ft/s','kts')'; % true airspeed, kts
KCAS = correctairspeed(KTAS,a,P,'TAS','CAS')'; % corrected airspeed, kts
qbar = dpressure([TAS zeros(size(TAS,1),2)], rho);
Dp = qbar*S.*C_D0;
Di = (2*W^2)/(rho*S*pi*e*A).*(TAS.^-2);
D = Dp + Di;
L = W;
h1 = figure;
plot(KCAS,L./D);
title('L/D vs. KCAS');
xlabel('KCAS'); ylabel('L/D');
hold on
plot(KCAS_bg,L_bg/D_bg,'Marker','o','MarkerFaceColor','black',...
    'MarkerEdgeColor','black','Color','white');
hold off
legend('L/D','L_{bg}/D_{bg}','Location','Best');
h2 = figure;
plot(KCAS,Dp,KCAS,Di,KCAS,D);
title('Parasite, induced, and total drag curves');
xlabel('KCAS'); ylabel('Drag, lbf');
hold on
plot(KCAS_bg,D_bg,'Marker','o','MarkerFaceColor','black',...
    'MarkerEdgeColor','black','Color','white');
hold off
legend('Parasite, D_p','Induced, D_i','Total, D','D_{bg}','Location','Best');
```